\documentclass[prd,aps,floats,epsfig,eqsecnum,nofootinbib]{revtex4}
\usepackage{amsmath,amssymb,verbatim,epsfig,rotating}
\def\be{\begin{equation}}
\def\ee{\end{equation}}
\newcommand{\bea}{\begin{eqnarray}}
\newcommand{\eea}{\end{eqnarray}}
\begin{document}
\title{Early Cosmology and Fundamental Physics\footnote{Based on
    Lectures at the 9th. Chalonge School in Astrofundamental Physics,
    Palermo, September 2002, NATO ASI. To appear in the Proceedings,
    N. S\'anchez and Yu. Parijskij editors, Kluwer.}}
\author{ {\bf H. J. de Vega} \\ \vspace{0.4cm}
LPTHE, Universit\'e
Pierre et Marie Curie (Paris VI) et Denis Diderot (Paris VII),
Tour 16, 1er. \'etage, 4, Place Jussieu, 75252 Paris, Cedex 05,
France}
\begin{abstract}
This is a pedagogical introduction to early cosmology and the host of 
fundamental physics involved in it (particle physics, grand unification and
general relativity). Inflation and the inflaton field are the central
theme of this review. The quantum field treatment of the inflaton is
presented including its out of equilibrium evolution and the use of
nonperturbative methods. The observational predictions for the CMB
anisotropies are briefly discussed.  Finally, open
problems and future perspectives in connection with dark energy and
string theory are overviewed.
\end{abstract}
\maketitle
\tableofcontents 

\section{The history of the universe}

The history of the universe is a history of expansion of the space and
cooling down. During all its history the universe as a whole is
homogeneous and isotropic in an excellent approximation, therefore it
is described by a Friedmann-Robertson-Walker (FRW) geometry 
\begin{equation}\label{FRW} 
ds^2= dt^2-a^2(t) \; d{\vec x}^2 
\end{equation}
where the scale factor $ a(t) $ grows with $t$. We consider the space
part flat according to the observed value of the density $ \Omega \simeq 1 $. 

Physical lengths increase as $ \times \; a(t) $ and the temperature
decreases as $ T(t) \sim \frac{1}{a(t)} $. 

Usually, time is parametrized by the redshift $z$ defined
according to astronomer's convention:
$$
1+z = \frac{a(\mbox{today})}{a(t)} \; .
$$
This formula gives the redshift of an event taking place at time $t$
in the past. Large $ z $ corresponds to early times when $  a(t) \ll
a(\mbox{today}) $.

A summary of the history of the universe is given in table 1.

\section{Fundamental Physics}

In order to describe the early universe we need:

\begin{itemize}
\item{General Relativity: Einstein's Theory of Gravity}

The matter distribution determines the geometry of the spacetime
through the Einstein equations. For the geometry eq.(\ref{FRW}), the
Einstein equations reduce to one scalar equation, the 
Einstein-Friedman equation
\be \label{ef}
\left[ \frac{1}{a(t)} \; \frac{da}{dt} \right]^2 = \frac{8 \, \pi}{3}
\; G \; \rho(t) \; ,
\ee
where $G$ stands for Newton's gravitational constant and $ \rho(t) $
for the energy density. 

\item{Quantum Field Theory and String Theory to describe Matter}

Since the energy scale in the early universe is so high (well beyond
the rest mass of particles), a quantum
field theoretical description for matter is unavoidable. Only such
context permits a correct description of particle production, particle
decays and transmutations. 

\end{itemize}

\bigskip

Electromagnetic, weak and strong interactions are well described 
by the so-called the standard model. That is, quantum
chromodynamics (QCD) combined with the electroweak theory
(electromagnetic and weak interactions). This a
non-abelian gauge theory associated to the symmetry group $ SU(3)
\otimes SU(2) \otimes U(1) $. The $SU(3)$ corresponding to the color
group of QCD while $ SU(2) \otimes U(1) $ describes the electroweak
sector. To this scheme, one adds presently neutrino masses (through the
see-saw mechanism) to explain  neutrino oscillations. 

The energy scale in QCD is about $ \sim 100$MeV $\simeq 10^{12}$K
corresponding to the chiral symmetry breaking while the energy scale
for the electroweak is the Fermi scale $\sim 100$GeV $\simeq
10^{15}$K. 

The standard model has been verified experimentally with spectacular
precision. However, it is an incomplete theory of particles. How to
complete it is a major challenge. It seems obvious that extensions of
the Standard model will be symmetric under a group containing $ SU(3)
\otimes SU(2) \otimes U(1) $ as a subgroup. The simplest proposals for
a Grand Unified Theory (GUT) include $ SU(5), \; SO(10), \; SU(6) $ and
$E_6$ as symmetry group.  

\begin{figure}[ht!]
\includegraphics[width=3.75in,keepaspectratio=true]{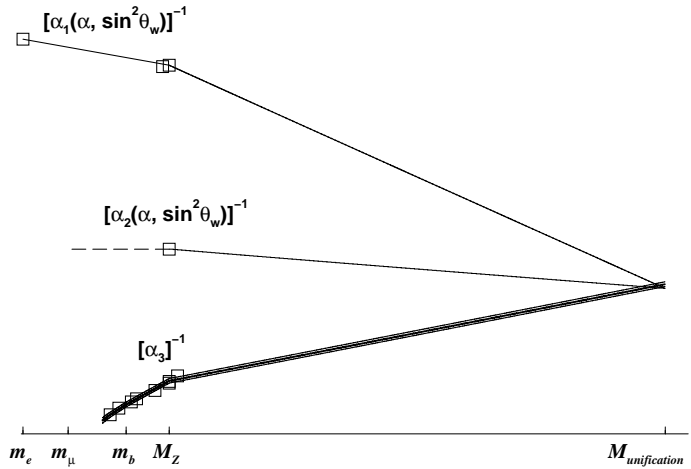}
\caption{Evolution of the weak $ \alpha_1 $, electromagnetic $
  \alpha_2 $ and strong $ \alpha_3 $ couplings with energy in the MSSM
  model. Notice that only  $ \alpha_3 $ decreases with energy
  (asymptotic freedom) extracted from ref.[9]} \label{unif}
\end{figure}

The  unification idea consists in that at some energy scale all three
couplings (electromagnetic, weak and strong) should become of the same
strength.  In this case, such grand unified scale turns out to be  $ E
\sim 10^{16}$GeV. 
The change of the couplings with the energy (or length) is governed in
physics by the renormalization group. 

Many extensions of the Standard model use supersymmetry in one way or
another. 

Supersymmetry transformations mix bosons and fermions. Supersymmetry
as well as GUT-symmetry should leave invariant the evolution laws (the
lagrangian) at sufficiently enough energy. Notice that the physical
states (or density matrices) describing the matter need not to be
invariant. For example, any thermal state at non-zero temperature
cannot be invariant under supersymmetry since Bose-Einstein and
Fermi-Dirac distributions are different. Somehow, the early universe
with $ T \gtrsim 10^{14}$GeV is one of the less supersymmetry
invariant situations in nature.

Generally speaking, the symmetry increases with energy. This is true
in general, in statistical mechanics, condensed matter as well as in
cosmology. For example, a ferromagnet at temperatures higher than the
Curie point is at the symmetric phase with zero magnetization. Below
the Curie point, the non-zero magnetization reduces the symmetry.

The same effect happens in the universe for symmetry breaking. The
universe started with maximal symmetry before inflation and this
symmetry reduces gradually while the universe expands and cools
off. The symmetry breaking transitions includes both the {\bf internal
symmetry groups} (as the GUT's symmetry group that eventually reduces
to the $ SU(3) \otimes SU(2) \otimes U(1) $ group) as well as the
translational and rotational symmetries which are broken by the density
fluctuations and the structure formation. These last produced by
gravity instabilities. 

It should be noticed, however, that no direct manifestation of
supersymmetry is known so far. An indication comes by studying the
energy running of the (electromagnetic, weak and strong) in the
standard model and in its minimal supersymmetric extension (MSSM). As
depicted in fig. \ref{unif} all three couplings meet at $ E \simeq  3
\times 10^{16}$GeV in the MSSM. The coupling unification becomes quite
loose in the Standard Model. This is why the renormalization group
running of the couplings in the MSSM supports the idea that
supersymmetry would a necessary ingredient of a GUT. For a recent
outlook see \cite{uno}.

It must be noticed that neutrino masses in the see-saw mechanism
naturally call for a mass scale of the order of the GUT scale. This is
the second evidence that an energy scale around $ 10^{15}$GeV plays a
crucial role. The third evidence comes from inflation (see below). 

\section{Essentials of Cosmology}

The essential observational evidence about the Universe can be
summarized as:

\begin{itemize}
\item{Isotropy and Homogeneity}

The Universe is isotropic and homogeneous for large enough
scales. Today this corresponds to $ L \gtrsim 100$Mpc. Galactic
surveys and the cosmic microwave background strongly support this
evidence. Therefore, the geometry of the universe is described by
eq.(\ref{FRW}). 

\item{Hubble Expansion}

Objects in the universe move at a speed proportional to their distance
according to Hubble's law:
$$
\frac{dR}{dt} = H(t) \; R(t) \quad \mbox{where} \quad H(t)=\frac{{\dot
    a}(t)}{a(t)} \; .
$$
This is an immediate consequence of eq.(\ref{FRW}). At present time $
H = 72 \pm 7 $km/(sec Mpc). Notice that objects at distances $ R >
\frac{1}{H} $ cannot be in causal contact with us since their velocity
would be higher than the velocity of light. $ d_H \equiv \frac{1}{H} $
is then the cosmological  horizon around us.

\item{Cosmic Microwave Background}

The CMB is isotropic up to $ \sim 10^{-4} $. It is the best known
black body spectrum with deviations less than $ 0.005 \%$. No
laboratory spectrum can beat it so far. Its temperature is $ T_{CMB} =
2.7277$K.

\end{itemize}

The Einstein-Friedman equation (\ref{ef}) is supplemented by the
continuity equation
\be \label{cont}
{\dot \rho}(t) + 3(t) H(t) \; \left[ \rho(t) + p(t) \right] = 0 \; ,
\ee
where  $p(t)$ stands for the pressure. The continuity equation follows
from the Einstein equations (Bianchi identity). 

The equation of state must be computed from the  appropriate theory of
matter considered. That is, the equation of state depends on the
nature of matter. 

\bigskip

\begin{tabular}{|c|c|c|c|}\hline
 &   &  & Era \\
Time   & Energy Scale  & Physical Phenomena & $1+z=a(\mbox{today})/a(t)$ \\ 
  & $1$ GeV $=1.16 \, 10^{13} $ K & & Scale Factor $a(t)$\\ \hline
 &   & Quantum Gravity  & $z>10^{26+20}=10^{46}$ \\ 
 $ \sim 10^{-44}$ sec. & $\sim10^{19}$GeV & String Theory & \\  
 & &Inflation starts & $a(t)\sim e^{Ht}$ \\ 
  & &  & Inflationary Era \\ 
 $ \sim 10^{-30}$sec. &$\sim10^{12}$GeV &Inflation Ends and & $z\sim
10^{20}$ \\ \hline 
 &   & Particle Creation Starts  &  \\
 &   & Reheating Transition  &  \\
 & & GUT Phase Transition & $a(t)\sim \sqrt{t}$ \\ 
 &   & Hot Big Bang: Thermalization &  \\
 &   &  &   \\
$ \sim 10^{-10}$sec. &  $\sim10^{3}$GeV & Electro-Weak Phase
Transition &   \\ 
&   & Baryon Asymmetry Originates? &  Radiation  \\
 &   &  &  \\
 &  $\sim 10^{2}$GeV  &Baryogenesis  & Dominated \\
 &   &  &  \\
 $ \sim 10^{-4}$sec. & $\sim 1$GeV   & Quark-hadron  and Chiral & Era \\
 $ \sim 10^{-2}$sec.&  $\sim 0.1$GeV  & Phase Transitions  &  \\
 &   &  &   \\
 &   &$\gamma, \; \nu, \; , e, \; {\bar e}, n , p$ in thermal
equilibrium  &  \\ 
 &   &  &   \\
 & $\sim 1$MeV  &Neutrinos decouple  &  \\
$1$sec. &   &  &  \\
 &   &Nucleosynthesis  &  \\
$100$sec. &  $\sim 0.1$MeV  & Creation of Light Elements  & $z\sim
10^4$ \\ \hline 
 &   &  &  \\
$20000$ years  &   & Structure Formation Begins & $a(t)\sim t^{2/3}$ \\
 &   &Onset of Gravitational Unstability  &  \\
 &   &  &   \\
$10^5$ years  &   &Atoms Form  & $z\sim 10^3$\\
 &   &  &   \\
 &   &  & Matter \\
 &   &Photon Decoupling  &Dominated  \\
 &   &The Universe Becomes Transparent  & Era  \\
 &   &  &  \\
$10^9$ years & first bound structures  &Galaxy Formation  &  \\
 &   &  & Cold matter dominates  \\
 &   &Solar system formation  &  but dark energy... \\
 &   &  &   \\
$1.4 \, 10^{10}$ years &  $\sim 10^{-4}$eV  & Today  & $z=1$ \\
 &   &  &   \\\hline
\end{tabular}

\vspace{0.5cm}

{TABLE 1. The history of the Universe. Time, typical energies and main
  physical phenomena from the begining till today}.

\bigskip

The most symmetric and simplest possibility is an energy-momentum
tensor proportional to the unit tensor,
$$
T_A^B = \Lambda \; \delta_A^B \quad 0\leq A, B \leq 3 \; .
$$
where $\Lambda$ is a constant. This corresponds to a constant energy
density $\rho(t)=\Lambda$ and a negative pressure $ p = - \rho $. 
$\Lambda$ is usually called the cosmological constant. 
A constant energy density in the Einstein-Friedman equation (\ref{ef})
leads  to an exponentially expanding universe with 
$$
a(t) = a(0) \; e^{H \; t} \; .
$$
This describes a De Sitter universe with $H = \sqrt{\frac{8\pi}{3} \; G
  \; \Lambda}$. 

Ultrarelativistic particles like radiation (hot matter) have the
equation of state $ p = \frac13 \; \rho$ which leads through
eqs.(\ref{ef}) and (\ref{cont}) to the FRW radiation dominated universe
with $ a(t) = \sqrt{t/t_0} $. The energy density dilutes here with
time as  $ \rho(t) = \rho_0 \; a(t)^{-4} $. 

Nonrelativistic particles (cold matter) obey $ p=0 $. It follows from
eqs.(\ref{ef}) and (\ref{cont}) the matter dominated universe with $
a(t) = \left(\frac{t}{t_0}\right)^{\frac23} $. The energy density
dilutes here with time as  $ \rho(t) = \rho_0 \; a(t)^{-3} $. That is,
they dilute the  rate of expansion of the space volume while massless
particles get an extra factor $a(t)$ due to the red-shift of their
energies. 

Strings exhibit a richer behaviour\cite{noscu}. Three different behaviours are
found with the following equations of state,
\begin{itemize}
\item{Stable strings} $ p = 0 $, they behave as nonrelativistic
  matter. 
\item{Unstable strings} $ p = - \frac{\rho}{3} $, this behaviour is
  only exhibited by strings. It implies a dilution
  $\rho=\frac{\rho_0}{a^2(t)} $. 
\item{Dual to unstable strings} $ p = \frac{\rho}{3} $, they behave as
  radiation. 
\end{itemize}
It must be stressed that a string during its time evolution {\bf
  changes } from one behaviour to another. A cosmological
  model describing the radiation dominated and matter dominated eras
  with strings is realized with a gas of classical strings\cite{noscu}. 

\section{Inflation and the Inflaton Field}

Inflation is part of the standard cosmology since several years. 

Inflation emerged in the 80's as the only way to explain the `bigness'
of the universe. That is, the value of the entropy of the
universe today $ \sim 10^{90} \sim (e^{69})^3 $. Closely related to
this, inflation solves the horizon and flatness problem explaining
therefore the quasi-isotropy of the CMB.  For a recent
outlook see \cite{dos}.

\bigskip

The inflationary era corresponds to the scale of energies of the Grand
Unification. It is not yet known which field model appropriately
describes the matter for such scales. Fortunately, one does not need a
detailed description in order to investigate inflationary cosmology. 
One needs the expectation value of the quantum energy density
($T_{00}$) which enters in the r. h. s. of the Einstein-Friedman
equation (\ref{ef}). This is dominated by  field condensates. Since
fermions fields have zero expectation values only the bosonic fields
are relevant. Bosonic fields do not need to be fundamental
fields. They can be pairs fermion-antifermion $ < {\bar \Psi} \Psi> $
in a GUT. In order to describe the cosmological evolution is enough to
consider the effective dynamics of such condensates. In fact, one
condensate field is enough to obtain inflation. It is usually called
`inflaton' and its dynamics can be described  by a Ginsburg-Landau
lagrangian in the cosmological background eq.(\ref{FRW}). That is, an
effective local Lagrangian containing terms of dimension less or equal
than four (renormalizable),
\be\label{lagra}
{\cal L} = a^3(t) \left[ \frac{{\dot \phi}^2}{2} - \frac{({\nabla
      \phi})^2}{2 \,  a^2(t)} - V(\phi) \right]
\ee
Here, the inflaton potential $ V(\phi) $ is usually a quartic
polynomial: $ V(\phi)= \frac{m^2}{2} \; \phi^2 + \frac{\lambda}{4}\;
\phi^4 $. 

The inflaton field $ \phi $ may not correspond to any real particle
(even unstable). It is just an effective description of the
dynamics. The detailed microscopical description should be given
by the GUT. Fortunately,  we do not need to know it in order to get a
description of the cosmological evolution. Indeed, a more precise
description should be possible from a microscopic GUT.
Somehow, the inflaton is to the  microscopic GUT theory
like the Ginzburg-Landau theory of superconductivity is to the
microscopic BCS theory. 

The inflaton model contains here two free parameters: $ m^2 $ and $
\lambda $. In order to reproduce the CMB anisotropies one has to
choose $ m^2 $ around the GUT scale and the coupling very small $
\lambda \sim 10^{-12}$. A model with only one field is clearly
unrealistic since the inflaton 
then describes a stable and ultra-heavy (GUT scale) particle. It is
necessary to couple the inflaton with lighter particles. Then, the inflaton
decays into them. 

\begin{figure}[htbp]
\rotatebox{-180}{\epsfig{file=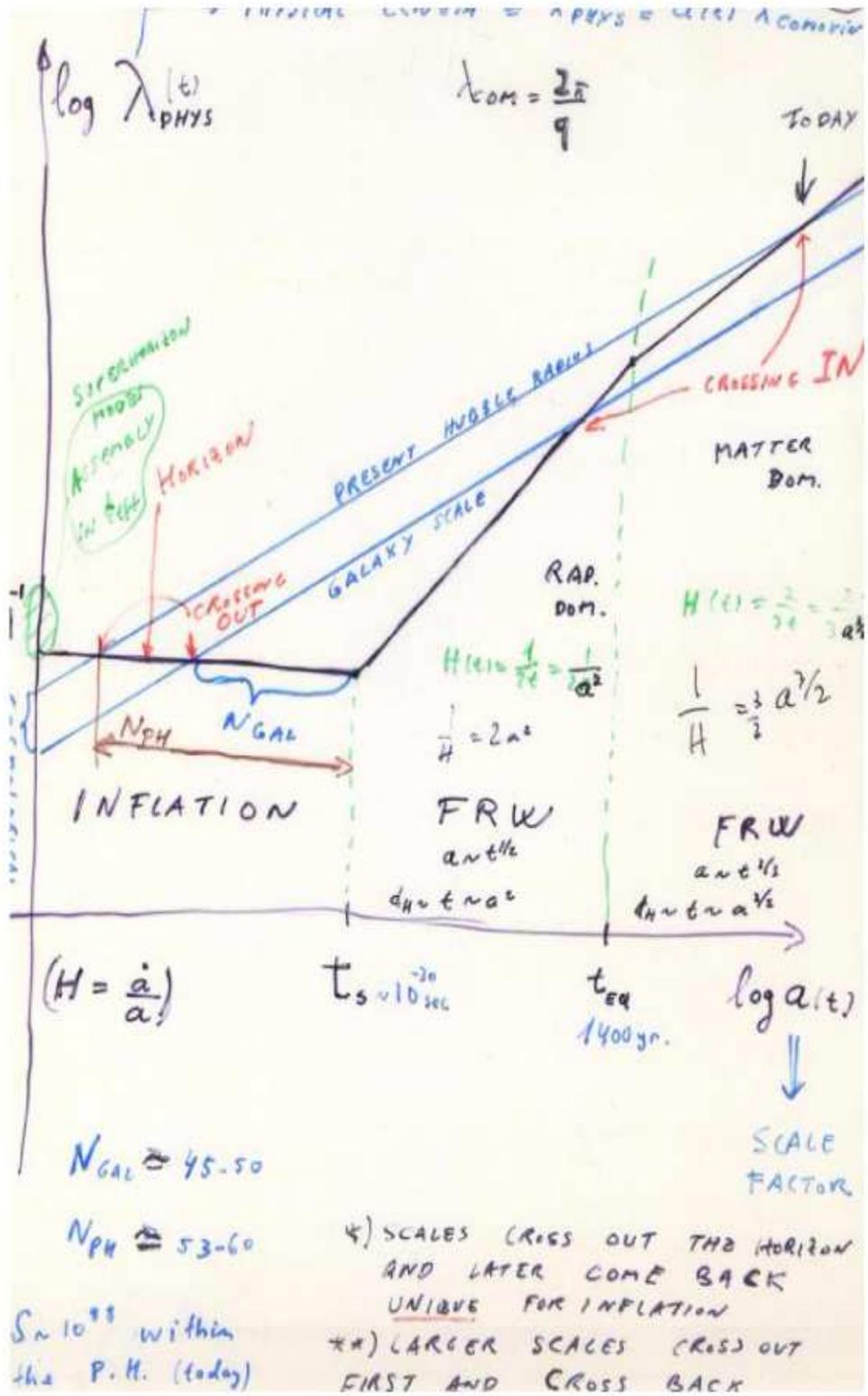,width=14cm,height=22cm}}
\caption{Physical lengths $ \lambda_{phys} = a(t) \;
  \lambda_{comoving} $ vs. the scale factor $  a(t) $ in a log-log
  plot. The black line is the horizon. One sees that some wavelengths
  can cross the horizon {\bf twice} bringing information from
  extremely short wavelength modes during the
  inflationary era. These transplanckian modes are responsible of the
  observed CMB anisotropies.} \label{flu}
\end{figure}

Fig. \ref{flu} shows how microscopic scales (even transplanckian) at the
begining of inflation become astronomical and produce  the CMB
anisotropies as well as the large scale structure of the universe. The
crucial fact is that the excitations can
cross the horizon {\bf twice}, coming back and bringing information
from the inflationary era. 

There are many available scenarios for inflation. Most of them add
extra fields coupled with the inflaton. This variety of inflationary
scenarii may seem confusing since many of them are compatible with the
observational data\cite{CMB}. Indeed, future observations should
constraint the models more tightly excluding some of them. Anyway, the
variety of acceptable inflationary models shows the power of the
inflationary paradigm. Whatever is the correct microscopic model for
the early universe, it must include inflation with the features we
know today.

The scenarii where the inflaton is treated classically are usually
characterized as small and large fields scenarii. In small fields
scenarii the initial classical amplitude of the inflaton is assumed
small compared with $|m| /\sqrt{\lambda}$ while in large field scenarii
the inflaton is initially of the order $\sim |m|
/\sqrt{\lambda}$\cite{lily}. The 
first type of scenarii is usually realized with broken symmetric
potentials ($ m^2 < 0 $) while for the second type scenarii (`chaotic
inflation') one can just use unbroken potentials with $ m^2 > 0 $.

\bigskip

Most of the work on inflation considers the inflaton field as a
classical field. This treatment is not accurate. The energy scales at which
inflation takes place call for a {\bf fully quantum} treatment of the
inflaton. This have been the subject of refs.\cite{cosm}-\cite{fond}. 

The coupled dynamics of the quantum inflaton and the geometry contains
rich physics. Therefore, we studied first the non-linear out of
equilibrium dynamics just in Minkowski spacetime\cite{mink}. Later,
the quantum inflaton dynamics in fixed cosmological backgrounds as de
Sitter\cite{fond}, radiation dominated and matter dominated
FRW. Finally the coupled dynamics of the inflaton and the scale factor
was studied in \cite{cosm} for a `new inflation' type scenario and in
ref.\cite{tsu} for tsunami inflation.  In all cases, out of
equilibrium field theory methods are used together with the
nonperturbative large $N$ approach in order to deal with the huge
 energy densities ($\sim m^4/\lambda$) non-analytic in $\lambda$.

In our treatment we consider gravity semiclassical: the geometry is
classical and the metric follows from the Einstein-Friedman equations
where the r.h.s. is the expectation value of a quantum
operator. Quantum gravity corrections can be neglected during
inflation because the energy scale of inflation $ \sim m_{inflaton}
\sim M_{GUT} \sim 10^{-5} \; M_{Planck} $. That is, quantum gravity
effects are at most $ \sim 10^{-5} $ and can be neglected in this
context. 

\section{Out of Equilibrium Field Theory and Early Cosmology}

As stressed before, the dynamics of inflation is essentially of 
quantum nature. This is extremely important, especially in light of
the fact that it is {\it exactly} this quantum behavior give rise to
the primordial metric perturbations which imprint the CMB. 

The inflaton must be treated as a {\it non-equilibrium} quantum field . The
simplest way to see this comes from the requirement of having small enough
metric perturbation amplitudes which in turn requires that the quartic self
coupling 
$ \lambda $ of the inflaton be extremely small, typically of order $ \sim
10^{-12} $. Such a small coupling cannot establish local thermodynamic
equilibrium (LTE) for {\it all} field modes; typically the long wavelength
modes will respond too slowly to be able to enter LTE. In fact, the
superhorizon sized modes will be out of the region of causal contact and
cannot thermalize. Out of equilibrium field theoretic  methods permit
us to follow the {\bf dynamics} of quantum fields 
in situations where the energy  
density is non-perturbatively large ($ \sim 1/\lambda $). That is, they allow
the computation of the time evolution of non-stationary states and of
non-thermal  density matrices.

Our programme on non-equilibrium dynamics of quantum field theory, started in
 1992\cite{cosm}-\cite{fond} to study the
 dynamics of  non-equilibrium processes from a fundamental
 field-theoretical description, by solving  the dynamical 
equations of motion of the underlying four dimensional quantum field
 theory for the early universe
 dynamics as well as  high energy particle collisions and phase
 transitions out of 
 equilibrium.

The focus of our work is to describe the quantum field dynamics when
the  energy density is {\bf high}. That is, a large number of particles per
volume $ m^{-3} $, where $ m $ is the typical mass scale in the
theory. Usual S-matrix calculations apply in the opposite limit of low
energy density and since they only provide information on {\em in}
$\rightarrow$ {\em out} matrix elements,  are unsuitable for calculations of
expectation values. 

In high  energy density situations such as in the early universe,
the particle propagator (or Green function) depends on the particle
distribution in momenta in a nontrivial way. This makes the 
quantum dynamics intrinsically nonlinear and calls to the use of
self-consistent non-perturbative approaches as the large $N$ limit.
In this approach, the inflaton becomes a $N$-component field 
$ \vec{\Phi} = (\Phi_1, \ldots , \Phi_N) $ with a selfcoupling of the
order $1/N$. That is, 
\begin{equation}
{\cal L} = a^3(t)\left[\frac{1}{2}\dot{\vec{\Phi}}^2(x)-\frac{1}{2} 
\frac{(\vec{\nabla}\vec{\Phi}(x))^2}{a^2(t)}-V(\vec{\Phi}(x))\right]
\label{action}
\end{equation}
\begin{equation}
V(\vec{\Phi})  =  \frac{m^2}2\; \vec{\Phi}^2 +
\frac{\lambda}{8N}\left(\vec{\Phi}^2\right)^2
+\frac12 \, \xi\; {\cal R} \;\vec{\Phi}^2  \;, \label{potential}
\end{equation}
where $ m^2 > 0 $ for unbroken symmetry and  $ m^2 < 0 $ for broken
symmetry. Here $ {\cal R}(t) $ stands for the scalar curvature
\begin{equation}
{\cal R}(t)  =  6\left(\frac{\ddot{a}(t)}{a(t)}+
\frac{\dot{a}^2(t)}{a^2(t)}\right), \label{ricciscalar}
\end{equation}
The $\xi$-coupling of $ \vec{\Phi(x)}^2 $ to the scalar curvature ${\cal
R}(t)$ has been included since  arises anyhow as a consequence of
renormalization\cite{fond}.

\bigskip

We consider translationally invariant quantum states $ |\Omega> $ or density
matrices $ {\hat \rho}(t) $ consistent with the
geometry. Inhomogeneities will arise as quantum fluctuations. That is,
$$
{\vec P} |\Omega> = 0 \quad \mbox{or}  \quad [{\vec P}, {\hat
    \rho}(t)] = 0 \; .
$$
The expectation value of the field $\Phi$ plays the role of order parameter
and can be chosen in a fixed direction (say 1) in the internal space,
$$
<\Phi_i ({\vec x},t)> \equiv\mbox{Tr}\left[  {\hat\rho}(t) \Phi_i
  ({\vec x},t) \right] = \delta_{i1} \; \phi(t)
$$
It is convenient to write the field operator  $\Phi$ as
$$
\Phi_i ({\vec x},t) = \delta_{i1} \;\phi(t) + \Psi_i ({\vec x},t)
\quad , \quad 1\leq i \leq N \; ,
$$
where the field operator  $\Psi$ has zero expectation value by
construction: $ <\Psi_i ({\vec x},t)> = 0 $.

Translational invariance allows to expand the field operator  $\Psi$ in
Fourier integral,
\be \label{campo}
\Psi_i ({\vec x},t) = \int \frac{d^3k}{(2\pi)^3} \left[ \alpha_i({\vec
    k}) \; \phi_k^*(t) \; e^{i {\vec k} \cdot {\vec x}} +
  \alpha_i^{\dagger}({\vec k}) \; \phi_k(t) \; e^{-i {\vec k} \cdot {\vec x}}
\right]
\ee
where the $  \alpha_i^{\dagger}({\vec k}), \; \alpha_i({\vec k}) , \;
1\leq i \leq N$ are creation-annihilation operators and the $ \phi_k(t) $ mode
functions. The  functions $ \phi_k(t) $ contain all the information
about the dynamics and are determined by the evolution equations and
the initial conditions as we
shall see below. For the vacuum state (Minkowski spacetime) the  $
\phi_k(t) $ have just a harmonic time dependence $ e^{i \, \omega_k \,
  t}$ with $ \omega_k = \sqrt{m^2 + k^2} $.

In the large $N$ limit this model becomes an infinite set of harmonic
oscillators with time dependent frequencies that contain the
expectation value $ < \left[ \Psi_i ({\vec x},t)\right]^2>$. That is,
the dynamics is non-linear and selfconsistent\cite{cosm,tsu}. 

The quantum fluctuations of $  \Psi_i $ are readily obtained from
eq.(\ref{campo}) with the result,
\be\label{Stau}
 < \left[ \Psi_i ({\vec x},t)\right]^2> =\int \frac{d^3k}{(2\pi)^3} \;
 |\phi_k(t)|^2 \; \coth\frac{\Theta_k}{2}
\ee
where the parameter $\Theta_k$ depends on the initial state chosen\cite{tsu}. 

In order to write the evolution equations it is convenient to choose
dimensionless variables:
\bea\label{dimvars1}
&&\tau = m \; t \quad ; \quad h(\tau)= \frac{H(t)}{m} \quad ;
\quad q=\frac{k}{m} \quad ;
\nonumber \\&&
\omega_q = \frac{\omega_k}{m} \quad ; \quad
\quad ; \quad g= \frac{\lambda}{8\pi^2}
\quad ; \quad
f_q(\tau) = \sqrt{m} \; \phi_k(t) \; ,\nonumber \\&&
g \; \Sigma(\tau) = \frac{\lambda}{2 \, m^2 \, N }  < \left[ \Psi_i ({\vec
x},t)\right]^2>  \quad ; \quad \eta(\tau) = \sqrt{\frac{\lambda}{2}}
\; \frac{\phi(t)}{m} \; .
\eea
The derivation of the equations of motion for the mode functions and
the field expectation value was given in ref.\cite{cosm,tsu} in the
limit $ N \to \infty $. The set of coupled, self-consistent
equations of motion for the quantum fields and the scale factor are
\begin{eqnarray}
&& \left[\frac{d^2}{d \tau^2}+3 \; h(\tau)
\frac{d}{d\tau}+{\cal M}^2(\tau) \right]\eta(\tau)  =  0 \cr \cr
&& \left[\frac{d^2}{d \tau^2}+3\; h(\tau)
\frac{d}{d\tau}+\frac{q^2}{a^2(\tau)}+{\cal M}^2(\tau)
\right]f_q(\tau)  =  0 \label{modknr}  \; .
\eea
Here, ${\cal M}^2(\tau) = \pm 1 + \eta^2(\tau)+g \; \Sigma(\tau)$ plays the
role of effective mass squared. The sign $+$ corresponds to unbroken
$O(N)$ symmetry while the sign $-$  corresponds to the broken symmetry
case. Notice that the zero mode (field expectation value) $ \eta(\tau)
$ obeys the same equation of motion as the $q=0$ mode.

The quantum fluctuations $ \Sigma(\tau)$ need to be subtracted
for ultraviolet divergences associated to mass and coupling constant
renormalization,
\be\label{Sig}
\Sigma(\tau)= \int_0^{\infty} q^2 dq \left[ | f_q(\tau)|^2 - \frac{1}{
{q\; a(\tau)^2}} + {\frac{\Theta(q - 1)}{2 q^3}} \left(\frac{{\cal
M}^2(\tau)}{m^2}-{\frac{{\cal{R}(\tau)}}{6 m^2}}\right)\right] \;
\ee
Notice that the  mass and coupling constant renormalizations are
identical to the Minkowski case since the high frequency regime is not
affected by the curved spacetime. 
Eqs.(\ref{modknr}) are coupled to the Einstein-Friedman equation for the scale
factor,
\begin{equation} \label{h2tau}
h^2(\tau) = L^2 \, \epsilon(\tau) \qquad ,\qquad
\mbox{where } L^2 \equiv \frac{16 \, \pi N \, m^2}{3\, M^2_{Pl}\, \lambda}
\; . 
\end{equation}
\noindent with the renormalized energy density $\epsilon(\tau)$ given
by\cite{tsu} 
\bea\label{renosubs}
\epsilon(\tau)  \equiv \frac{\lambda}{2 N \; m^4} \langle
T^{00}\rangle_R &=& \frac{g\Sigma(\tau)}{2} + \frac{[g\Sigma(\tau)]^2}{4} +
\frac{g}{2}\int q^2 \; dq \left\{|\dot{f_q}(\tau)|^2 -
S_1(q,\tau)+ \right. \cr\cr
&+&\left.\frac{q^2}{a^2(\tau)} \left[|f_q(\tau)|^2 - S_2(q,\tau)\right]
\right\} \; , 
\eea
\bea \nonumber
&&S_1(q,\tau) =\frac{q}{a^4(\tau)}+\frac{1}{2qa^4(\tau)}
\left[B(\tau)+2\dot{a}^2  \right] 
+ \frac{\Theta(q - 1)}{8 q^3 \; a^4(\tau)} \left[ - B(\tau)^2
- a(\tau)^2 {\ddot B}(\tau) +\right. \cr\cr
&&\left. +3 \, a(\tau) \,  {\dot a}(\tau) \,
{\dot B}(\tau) - 4 {\dot a}^2(\tau) \, B(\tau) \right]\;,\cr\cr
&&S_2(q,\tau) = \frac{1}{qa^2(\tau)}- \frac{1}{2q^3 a^2(\tau)}\;B(\tau)
+ \frac{\Theta(q - 1)}{8 q^5 \; a^2(\tau) }\left\{  3 B(\tau)^2
+ a(\tau) \frac{d}{d\tau} \left[ a(\tau) {\dot B}(\tau)\right]
\right\} \; , \cr\cr
&&\mbox{with} \; B(\tau) \equiv a^2(\tau)\;{\cal M}^2(\tau) \; .\nonumber
\end{eqnarray}
The subtractions performed essentially correspond to the divergent
part of the zero point fluctuations and ensure the finiteness of the
energy density as well as the covariant conservation (Bianchi identity)
$$
{\dot \epsilon}(\tau) + 3 \;  h(\tau) \left[ p(\tau) + \epsilon(\tau)
  \right] = 0 \; ,
$$
where the renormalized pressure $ p(\tau) $ follows from
eq.(\ref{renosubs}) and \cite{tsu} 
\be\label{pres}
p(\tau)+\epsilon(\tau)= g \int q^2 dq \left\{ |\dot{f_q}(\tau)|^2 - S_1(q,\tau)
+\frac{q^2}{3a^2(\tau)}\left[  |f_q(\tau)|^2 - S_2(q,\tau) \right] \right\}
\;.
\ee
These evolution equations are supplemented by the initial
conditions. In general, they take the form,
\bea\label{omegaq}
&& \eta(0) = \eta_0 \quad , \quad {\dot \eta}(0) = \xi_0 \quad , \quad
f_q(0)  =  \frac{1}{\sqrt{\Omega_q}} \quad , \quad
\dot{f}_q(0)  = - \left[\omega_q \; \delta_q + h(0)
+ i\,\Omega_q \right]f_q(0) \label{condini} \nonumber \\
&&\omega_q = \sqrt{q^2+\left|1+\eta_0^2 + g\Sigma(0)-\frac{{\cal
R}(0)}{6m^2}\right|} \quad , \quad a(0) = a_0 \quad , \quad h(0) = h_0 \; . 
\end{eqnarray}
We usually choose $  \xi_0 = 0 $ since one can always enforce that by
a shift in the time variable. For simplicity we take $ a_0 = 1 $. The
parameters $ \Omega_q $ and $ \delta_q $ are arbitrary and
characterize the initial density matrix (the initial state). 

The inflationary scenarii described by a classical inflaton field
correspond to the choice of the vacuum state for the oscillators. That
is, $ \Omega_q = \omega_q , \; \delta_q = 0 $. In that case the
quantum fluctuations are of the order one  at the initial time
$\Sigma(0)= {\cal O}(1) $ and hence $g \;\Sigma(0) \ll 1 $.

The tsunami inflationary scenario correspond to a band
of  excited quantum modes in the initial state, thus the name
`tsunami-wave'\cite{tsu}. This initial state
models a cosmological initial condition in which the energy density
is non-perturbatively large, but concentrated in the
quanta rather than in the field expectation value.

In summary, we have an infinite number of coupled non-linear
differential equations (\ref{modknr})-(\ref{renosubs}) which are {\bf
  local} in time but non-local in the wavenumbers $q$. The unknowns
are $ a(\tau), \; \eta(\tau), \; f_q(\tau), \; 0 \leq q \leq \infty $ and
the initial conditions are listed in eq.(\ref{omegaq}).

Eqs.(\ref{modknr})-(\ref{renosubs}) can be solved analytically for
short and late times. Otherwise, the numerical treatment is easy to
implement. Notice that all physical quantities are computed from the
mode functions $f_q(\tau), \; 0 \leq q \leq \infty$, the zero mode
$\eta(\tau)$  and the scale factor $ a(\tau) $. For example, the
equal-time correlators of the field $\Psi$ are given by
$$
<\Psi({\vec x},\tau) \; \Psi({\vec y},\tau)> = \int \frac{d^3q}{(2\pi)^3} \;
 |f_q(\tau)|^2 \; \coth\frac{\Theta_q}{2} \; .
$$
The resolution of eqs.(\ref{modknr})-(\ref{renosubs}) is discussed in
detail in refs.\cite{cosm} as well as the observational implications
through the density fluctuations. We present here now the crucial
features. 

Let us consider a new inflation scenario with broken symmetry as in
ref.\cite{cosm}. Assuming initially the ground state for the  quantum
modes $ f_k $ and  $  \eta_0 = \xi_0 = 0 $ (opposite to the tsunami
scenario \cite{tsu}). We then have that  $g \;\Sigma(0) \ll 1 $, and
we can approximate the effective mass for short times as  $ {\cal
  M}^2(\tau) =  - $ in  the mode equations (\ref{modknr})
$$
\left[\frac{d^2}{d \tau^2}+3\; h_0 
\frac{d}{d\tau}+\frac{q^2}{[a_0]^2 \; e^{2 \; h_0 \; \tau}}-1
\right]f_q(\tau)  =  0
$$
with solution 
\begin{equation}\label{bessel}
f_q(\tau) = e^{-\frac{3}{2} \; h_0 \;  \tau} \left[ a(q)\;
  J_{\nu}\left(\frac{q}{h} \; e^{-h_0 \; \tau}\right)+b(q)\; 
J_{-\nu}\left(\frac{q}{h} \; e^{-h_0 \; \tau}\right) \right] \; ,
\end{equation}
where $ \nu \equiv  \sqrt{\frac{1}{h^2}+\frac{9}{4}} \; , J_{\nu}(z) $
stand for a Bessel function and the
coefficients $a(q)$ and $b(q)$ are determined by the initial
conditions\cite{cosm}. Since the argument of the Bessel functions
tends to zero very fast for increasing time, the second term in
eq.(\ref{bessel}) dominates and we have
\begin{equation}\label{Uasi}
f_q(\tau) \simeq e^{-\frac{3}{2}\; h_0 \;  \tau} \; b(q) \;
J_{-\nu}(\frac{q}{h} \; e^{-h_0 \; \tau}) \simeq \frac{b(q)} 
{\Gamma(1-\nu)} \; \left( \frac{2h}{q}\right)^{\nu} \; e^{(\nu-3/2)h_0\;
  \tau}   \;. 
\end{equation}
We see that the mode functions {\bf grow} exponentially fast in time
due to the spinodal instabilities. [Recall that broken symmetry
  implies $ m^2 < 0 $ in  $ V(\phi) $ eq.(\ref{lagra})].
Notice that the growth of the modes here (de Sitter spacetime) is
{\bf different} to the  growth in Minkowski and FRW
spacetimes\cite{mink,fond}. 
Notice furthermore that eq.(\ref{Uasi}) applies when $
\lambda_{phys} = 2 \pi \; e^{h_0 \; \tau} / q >  2 \pi/h_0 $. That is,
after the modes have crossed out the horizon. The physical meaning of
the growth of the mode functions is that particles are created at the
expense of the dark energy (uniformly distributed energy) driving inflation.

\bigskip

This linear approximation breaks down when $ g \Sigma(\tau) $ is no
more negligible compared with unity. Estimating $ \Sigma(\tau) $ for
short times from eqs.(\ref{Stau}) and (\ref{Uasi}) yields \cite{cosm},
\begin{equation}\label{taunl}
\tau_{nl} \simeq \frac{1}{(2\nu-3) \; h_0}\ln\frac{1}{g} + {\cal O}(1)
\ee
For $\tau > \tau_{nl}$ (the nonlinear time), the nonlinear effects of
backreaction through  $ \Sigma(\tau) $ become very important, and the 
contribution from the quantum fluctuations competes with the tree level terms
in the equations of motion, shutting-off the instabilities. Beyond $\tau_{nl}$,
the full numerical analysis of eqs.(\ref{modknr})-(\ref{renosubs})
shown in figs. \ref{gsigma}--\ref{modu} captures the correct dynamics. 

Figs. \ref{gsigma}--\ref{modu} show $g\Sigma(\tau)$ vs. $\tau$,
$h(\tau)$ vs. $\tau$ and $ \ln|f_q(\tau)|^2 $ vs. $\tau$ for several values
of $q$ with larger $q's$ corresponding to successively lower curves. 
They correspond to $g = 10^{-14} \; ; \; \eta_0=0 \; ; \; \xi_0=0$
and we have chosen the representative value $h_0=2.0$.

Figs. \ref{gsigma} and \ref{hubblefig} show clearly that 
when the contribution of the quantum
fluctuations $ g\Sigma(\tau) $ becomes of order 1 inflation ends,
and the time scale for $ g\Sigma(\tau) $ to reach ${\cal O}(1)$ is very well
described by  the estimate eq.(\ref{taunl}). From
fig.\ref{gsigma} we see that this happens for $ \tau =\tau_{nl} \approx
90 $, leading to a number of e-folds $ N_e \approx 180 $ which is
correctly estimated by eq. (\ref{taunl}).

\begin{figure}
\epsfig{file=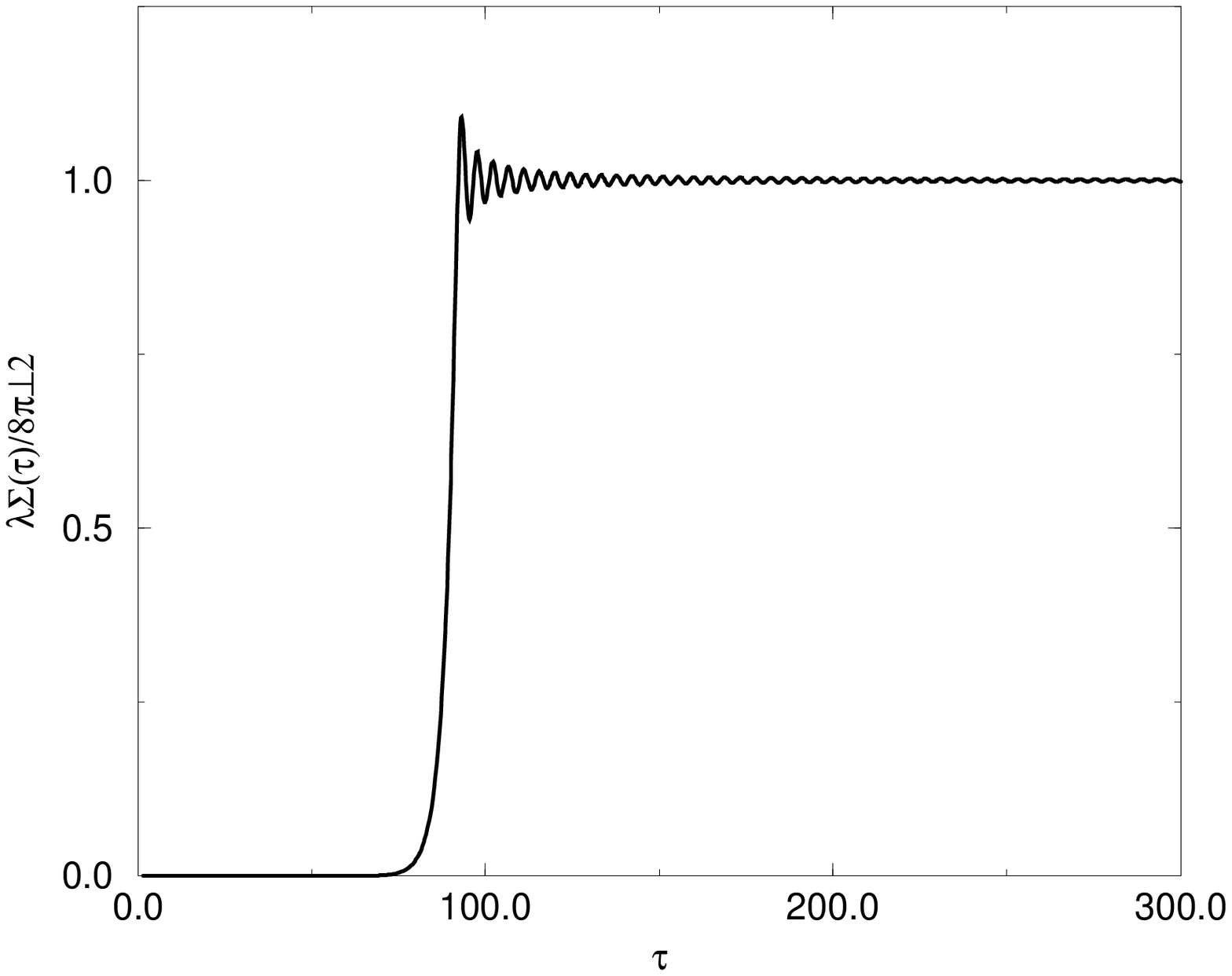,width=12.5cm,height=8.5cm}
\caption{ $ g\Sigma(\tau) $ vs. $\tau$, for $\eta(0)=0, \dot{\eta}(0)=0,\, 
g = 10^{-14},\,  h_0 = 2.0$. }
\label{gsigma}
\epsfig{file=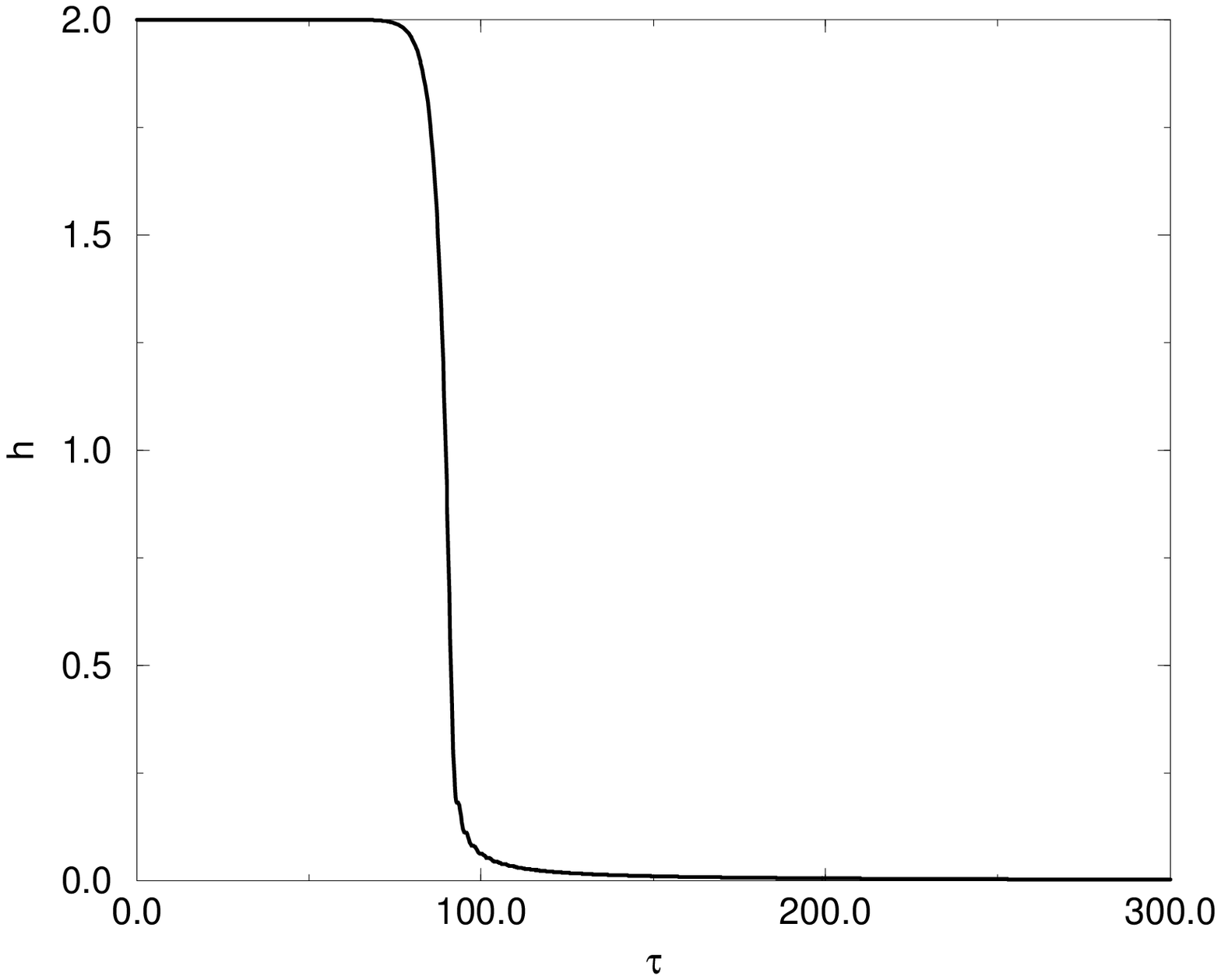,width=12.5cm,height=8.5cm}
\caption{$ h(\tau) $ vs. $ \tau $, for $ \eta(0)=0,\,  \dot{\eta}(0)=0, \, 
g = 10^{-14},\,  h_0 = 2.0 $. }
\label{hubblefig}
\end{figure}

Fig. \ref{modu} shows clearly the factorization of the modes after they
cross the horizon as described by eq.(\ref{Uasi}).
 The slopes of all the curves after they become
straight lines in fig.\ref{modu} is given exactly by $ 2\nu-3 $,
whereas the 
intercept depends on the initial condition on the mode function and
the larger the value of $ q $ the smaller the intercept because the
amplitude of the mode function is smaller initially. 
 Notice from the figure that when inflation ends and
the non-linearities become important all of the modes effectively saturate.
This is also what happens for  the zero mode:
exponential growth in early-intermediate times (neglecting the
decaying solution), with a growth exponent
given by $ \nu - 3/2 $ and an asymptotic behavior of small oscillations
around the equilibrium position, which for the zero mode is $\eta =1$, but
for the $q \neq 0$ modes depends on the initial conditions. 
All of the mode functions have this behavior once they cross the horizon.
We have also studied the phases of the mode functions and we found that 
they freeze after horizon crossing in the sense that they become independent
of time. This is natural since both the
real and imaginary parts of $ f_q(\tau) $ obey the same equation but
with different 
boundary conditions. After the physical wavelength crosses the horizon, the
dynamics is insensitive to the value of $q$ for real and imaginary parts and
the phases become independent of time. Again, this is a consequence of the
factorization of the modes. 

\begin{figure}
\epsfig{file=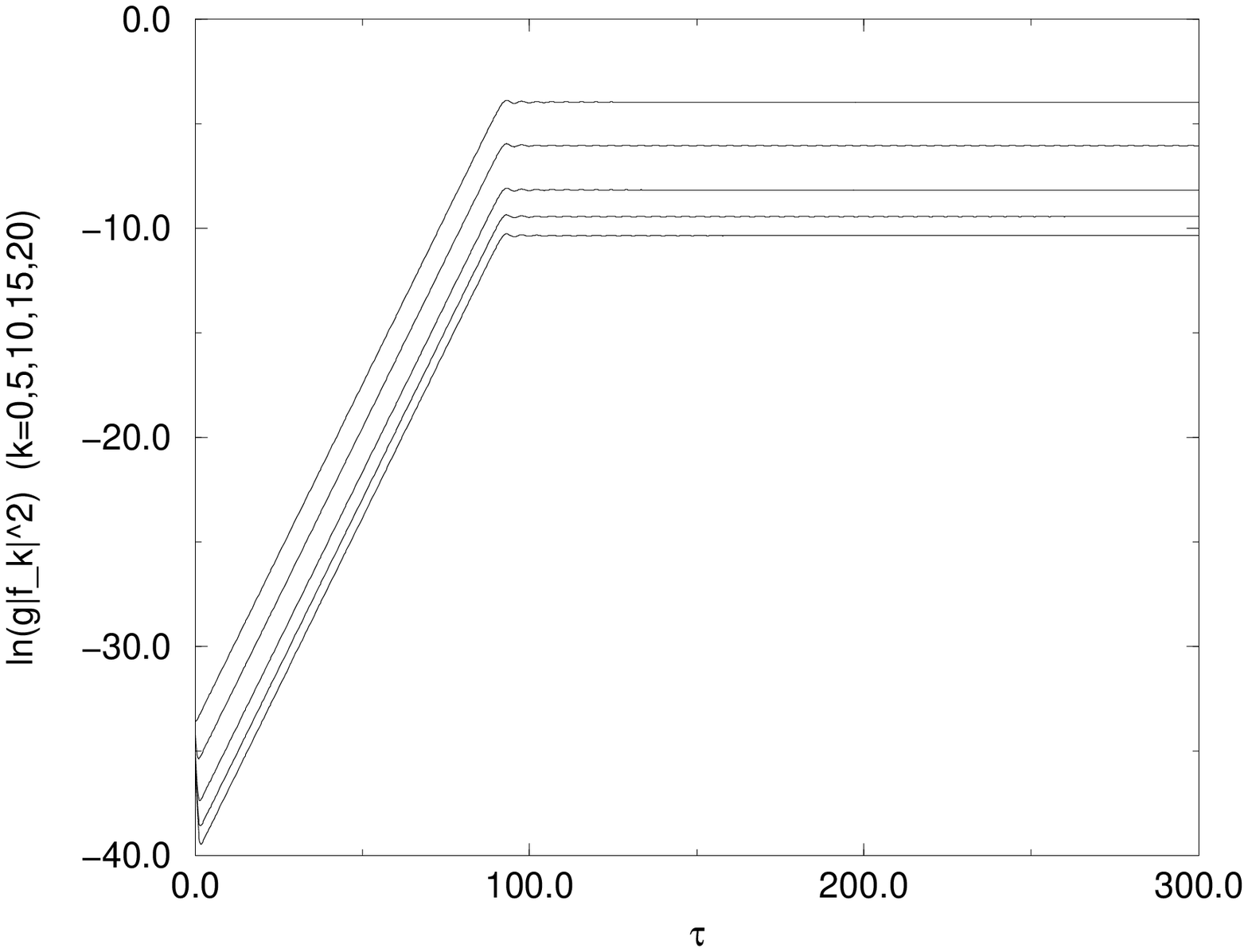,width=12.5cm,height=8.5cm}
\caption{$ \ln|f_q(\tau)|^2 $ vs. $\tau$, for $\eta(0)=0,
\dot{\eta}(0)=0,  g = 10^{-14}, h_0=2.0$ for
$q=0.0,5,10,15,20$ with smaller 
$q$ corresponding to larger values of $\ln|f_q(\tau)|^2 $.}
\label{modu}
\end{figure}
The growth of the quantum fluctuations ends inflation
at a time given by $\tau_{nl}$ [eq.(\ref{taunl})]. Furthermore,
the calculation of the pressure from eq.(\ref{pres}) shows that during
the inflationary epoch $p(\tau)/\varepsilon(\tau)  
\approx -1$ and the end of inflation is rather sharp at $\tau_{nl}$.
$p(\tau)/\varepsilon(\tau)$ oscillates between $\pm 1$ with zero average
over the cycles, resulting in matter domination. Moreover,
$h(\tau)$ is constant (and equals to $ h_0 $) during the de Sitter epoch and
becomes matter dominated after the end of inflation with $h^{-1}(\tau) 
\approx \frac32 (\tau -\tau_{nl})$. 

All of these features hold for a variety of initial conditions.  As an
example, we show in ref.\cite{cosm} the case of an initial Hubble
parameter of $h_0=10$. The reason why our results are independent on
the details of the initial conditions stemmed from the fact that the
spinodal instabilities dominate the dynamics. Therefore, small changes
on the initial data only have an  irrelevant physical effect. The same
is true for tsunami inflation\cite{tsu}.

\bigskip

We computed in ref.\cite{cosm} the density fluctuations for
cosmologically relevant modes (see fig. 2). We found,
\begin{equation}\label{amplitude}
|\delta_k(t_f)| = \frac{3}{ 5 \; \pi} \frac{ \Gamma(\nu)}{
(\nu-\frac{3}{2})\,  {\cal F}(H_0/m)}
\left(\frac{2H_0}{k}\right)^{\nu-\frac{3}{2}} \; , 
\end{equation}
where the function ${\cal F}(H_0/m)$ computed in in ref.\cite{cosm}
encodes the information from the quantum fluctuations $\Sigma(\tau)$.
We read the power spectrum per logarithmic $k$ interval as,
\begin{equation}
{\cal P}_s(k) = |\delta_k|^2 \propto k^{-2(\nu-\frac{3}{2})}.
\end{equation}
leading to the index for scalar density perturbations
\begin{equation}
n_s = 1-2\left(\nu-\frac{3}{2}\right) \; . \label{index}
\end{equation}
The recent WMAP observations\cite{CMB} are compatible with this {\bf
  red} spectrum. 
\section{Open Problems and Outlook}

The Universe today is formed by a $73\%\pm 4\% $ of dark energy and
$23\%\pm 4\% $ of dark matter. Here are two of the greatest open
questions. 

Dark energy is continuously distributed energy in the Universe. It
amounts to a uniform density of about four proton masses per cubic meter. 
It seems natural to think that it is due to zero point quantum
fluctuations. A naive calculation in a fixed spacetime yields a
completely wrong order of magnitude. It must then be a {\bf dynamical}
quantity that evolves with the universe. No theoretical explanation
for the dark energy is available today. Notice, however that the dark
energy in the inflationary universe plays a clear role in a quantum
field theory treatment and transforms into the created particles (see
sec. V and ref.\cite{cosm,tsu}).
Dark matter is a further open problem. The  nature of the particles
that forms it is still unknown. 

The physics beyond GUT's is a fascinating but unchartered territory.
The quantum gravity phenomena receive a great deal of attention since
many years\cite{string}. It may be very
 well that a quantum theory of gravitation needs new concepts and ideas.
 Of course, this future theory must have the today's General Relativity and
 Quantum Mechanics (and QFT) as limiting cases. In some sense, what everybody
  is doing in this domain (including string theories approach)
may be something analogous to the development of the old quantum theory
in the 10's of this century\cite{streri}. Namely, people at that time
imposed quantization conditions (the Bohr-Sommerfeld conditions) to
hamiltonian mechanics but keeping the  concepts of  classical mechanics.

The main drawback to develop a quantum theory of gravitation is
 clearly the {\bf total lack of experimental guides} for the theoretical
 development. Just from  dimensional reasons, physical effects combining
 gravitation and quantum mechanics are relevant only at energies of the
 order of  $M_{Planck}  =  \sqrt{ \hbar c / G } =  1.22 \,  10^{16} $Tev.
 Such energies were available in the Universe at times $ t \sim t_{Planck}
 =  5.4 \, 10^{-44} $sec. Anyway, as a question of principle,
 the construction of a quantum theory of gravitation
 is a problem of fundamental  relevance for physics and cosmology.

Recent work pays a great deal of attention to branes in string
theory. These are {\bf classical} vacua of string theory. The quantum vacuum
of string theory is unfortunately unknown. It is not yet clear whether
branes teach us anything about it. Notice that only in the simplest
field theories (like $\phi^4$) the classical and the quantum vacuum
are the same. Already in Yang-Mills theory or QCD, the classical and
quantum ground state are radically different.

\end{document}